\documentclass[11pt,fleqn]{ndrpt}
\usepackage{ndthms}

\usepackage{amssymb}
\usepackage{hyperref}

\title{The weak pigeonhole principle for function classes in $S^1_2$}
\makeatletter
\def\subjclassname@ACM{ACM Subject Classification}
\makeatother

\author{Norman Danner${}^*$}
\address{%
Department of Mathematics and Computer Science \\
Wesleyan University \\
Middletown, CT 06549}
\email{ndanner@wesleyan.edu}
\thanks{${}^*$Corresponding author}

\author{Chris Pollett}
\address{%
214 MacQuarrie Hall \\
Department of Computer Science \\
San Jose State University \\
One Washington Square, San Jose CA 95192}
\email{pollett@cs.sjsu.edu}

\theoremstyle{ndproclaim}
\newtheorem{lem}[thm]{Lemma}
\newcommand{\Hajek}{H\'{a}jek}
\newcommand{\Pudlak}{Pudl\'{a}k}
\newcommand{\Krajicek}{Kraj\'\i\v{c}ek}
\newcommand{\Jerabek}{Je\v{r}\'{a}bek}

\newcommand{\compfont}{\mathsf}

\newcommand{\SIG}[1]{\Sigma^{\compfont b}_{#1}}
\newcommand{\PI}[1]{\Pi^{\compfont b}_{#1}}
\newcommand{\DELT}[1]{\Delta^b_{#1}}

\newcommand{\NP}{\compfont{NP}}

\newcommand{\mathfnfont}{\mathit}
\newcommand{\mathpredfont}{\mathit}

\def\lh#1{\left|#1\right|}
\def\lhlh#1{\left\|{#1}\right\|}

\newcommand{\proj}[2]{\beta(#1,#2)}
\newcommand{\SqBd}{\mathfnfont{SqBd}}

\newcommand{\PAD}{\mathfnfont{PAD}}

\newcommand{\MSP}{\mathfnfont{MSP}}
\newcommand{\BIT}{\mathfnfont{BIT}}
\newcommand{\Bit}[1]{\mathfnfont{BIT}(#1)}
\newcommand{\Seq}{\mathfnfont{Seq}}
\newcommand{\proves}{\vdash}
\newcommand{\monus}{\mathbin{\mathchoice%
{\buildrel .\lower.6ex\hbox{\vphantom{.}} \over {\smash-}}%
{\buildrel .\lower.6ex\hbox{\vphantom{.}} \over {\smash-}}%
{\buildrel .\lower.4ex\hbox{\vphantom{.}} \over {\smash-}}%
{\buildrel .\lower.3ex\hbox{\vphantom{.}} \over {\smash-}}}}
\newcommand{\IFF}{\Leftrightarrow}
\newcommand{\AND}{\wedge}
\renewcommand{\AND}{\mathrel{\land}}
\newcommand{\OR}{\vee}
\renewcommand{\OR}{\mathrel{\lor}}
\newcommand{\NOT}{\neg}
\newcommand{\IMP}{\rightarrow}

\newcommand{\HALF}[1]{\lfloor\frac{1}{2}#1\rfloor}  
\newcommand{\nat}{\mathbb{N}}
\newcommand{\theoryfont}{\mathit}

\newcommand{\BASIC}{\theoryfont{BASIC}}
\newcommand{\TT}[1]{\theoryfont{T}^{#1}_2}
\newcommand{\ST}[1]{\theoryfont{S}^{#1}_2}

\newcommand{\LIND}{\theoryfont{LIND}}
\newcommand{\IND}{\theoryfont{IND}}
\newcommand{\WPHP}{\theoryfont{WPHP}}
\newcommand{\PHP}{\theoryfont{PHP}}
\newcommand{\REPL}{\theoryfont{REPL}}

\newcommand{\COMP}{\theoryfont{COMP}}

\newcommand{\BPR}{\theoryfont{BPR}}

\newcommand{\BPS}{\mathfnfont{PfxSeries}}
\def\BS{\mathpredfont{BitSeries}}
\let\PS\BPS
\def\BSBOUND#1#2#3{\mathpredfont{BitBound}_{#1,#2,#3}}
\def\PSBOUND#1#2#3{\mathpredfont{PfxBound}_{#1,#2,#3}}

\newcommand{\eval}{\mathop{\mathfnfont{eval}}}
\newcommand{\floor}[1]{\lfloor #1\rfloor}

\newcommand{\id}{\mathop{\mathrm{id}}}

\def\PSqBd{\mathfnfont{PSqBd}}
\def\pair#1#2{\langle#1,#2\rangle}

\def\setseparator{:}
\newcommand{\seq}[2][\relax]{
  \ifx#1\relax
    \langle#2\rangle
  \else
    \ifx#1\left
      \left\langle#2\right\rangle
    \else
      \csname #1l\endcsname\langle#2\csname #1r\endcsname\rangle
    \fi
  \fi
}

\newcommand{\set}[2][\relax]{
  \ifx#1\relax
    \{#2\}
  \else
    \ifx#1\left
      \left\{#2\right\}
    \else
      \csname #1l\endcsname\{#2\csname #1r\endcsname\}
    \fi
  \fi
}
\newcommand{\setst}[3][\relax]{\set[#1]{#2\setseparator#3}}

\begin{document}

\begin{abstract}
It is well known that $\ST1$ cannot prove the injective weak pigeonhole principle
for polynomial time functions unless RSA is insecure. In this note we investigate
the provability of the surjective (dual) weak pigeonhole principle in $\ST1$ for provably weaker
function classes.
\end{abstract}

\maketitle

\section{Introduction}
The weak pigeonhole principle for a relation $R(x,y)$ says that $R$ does not represent an injective map from
$n^2$ pigeons to $n$ holes. 
Variants of the weak pigeonhole principle have been shown to be connected with cryptography 
and circuit lower bounds in several different ways. \Krajicek \,and \Pudlak~\cite{krajicek-pudlak:ic98}  have shown that 
if the theory $\ST{1}$ can prove the principle for graphs of $p$-time functions then the cryptographic scheme RSA is insecure. Here $\ST1$ is roughly a theory which has axioms for the symbols of arithmetic and length induction axioms for  $\NP$-predicates. The surjective (dual)
variant of the weak pigeonhole principle states there is no surjective
map from $n$ pigeons onto $n^2$ holes.%
\footnote{Some authors refer to the principle that
asserts that there is no bijective map from~$n^2$ pigeons onto
$n$ holes as the \emph{onto} principle; we shall not refer to this
principle or use this terminology in this paper.}
\Jerabek~\cite[\S3]{jerabek:apal04} has shown that the surjective weak pigeonhole principle for $p$-time functions is equivalent over~$\ST1$ to 
(essentially) the 
schema that asserts that for each fixed~$k>0$ that there is a string
of length~$2n^k$ that cannot be bit-recognized by any circuit of size~$n^k$.
More recently, Pollett and Danner~\cite{pollett-danner:tcs06} 
have shown that the multifunction weak pigeonhole principle for iterated $p$-time relations is equivalent over $\ST1$ to the existence of strings 
that are hard for an iterated circuit block  
 recognition principle. This implies that if RSA is secure then $\ST1$ cannot prove superpolynomial circuit lower bounds for multifunctions computed by iterated $p$-time relations. In an attempt to make progress towards making these contingent results non-contingent, the present note investigates whether there are any interesting classes of functions for which $\ST1$ \emph{can} prove the weak pigeonhole principle.
 
Proofs of the pigeonhole principle usually start by assuming one has a map that violates the pigeonhole principle, then constructing a submap that also violates the pigeonhole principle and applying induction to get an obvious
contradiction, such as an injective map of two objects into one. 
The weakest theory known to prove the weak pigeonhole principle for 
graphs of $p$-time multifunctions is $\TT{2}$, which is defined like $\ST1$ but with usual induction for $\NP^{\NP}$-predicates. This was shown by 
Maciel et al.~\cite[\S6]{maciel-etc:wphp}  following essentially this paradigm. The authors assume that they have a  multifunction mapping $n^2$ pigeons to $n$ holes. 
The pigeons are split into groups of size~$n$ and the
holes into two groups of size $n/2$. 
They then argue that either (1)~all of one group of pigeons must be mapped into the first group of holes, or (2)~one can pick one pigeon from each group so that pigeons from different groups are mapped to different holes (all in
the second group).  
In either case one gets a map from $n$ pigeons to $n/2$ holes
which is amplified to a map from~$n^2$ pigeons
to $n/2$ holes using the original map.  This process is then iterated.
The entire argument is carried out in~$\ST3$, which is conservative
over~$\TT2$ for $\SIG3$ formulas (an in particular, for the
weak pigeonhole principle).
 
In contrast to the above technique for proving the weak pigeonhole principle, in the current paper we use a technique that clearly illustrates the cryptographic nature of these principles. We consider a function algebra $A^3$ which is the closure of the terms of the language of $\ST1$ under $3$-lengths bounded primitive recursion (see Definition~\ref{defn:bpr}). 
Working in~$\ST1$ we show that any function in~$A^3$ omits values
of the form~$\floor{(n\#n-1)/3}$ from its range, where
$x\#y = 2^{\lh x\lh y}$.
Pollett~\cite{pollett:mfn-algebras,pollett:aml03} has connected
the algebras~$A^m$ to weak theories of arithmetic, and the techniques
of those papers can be used to show that if $f(x)\in A^4$, 
then $f(x)\not=\lfloor x/3 \rfloor$.
In this paper we prove the much harder statement that for any~$n$ and
any $a\leq n$, $f(a)\not=\floor{(n\#n-1)/3}$; in particular, $f(\vec x)$
is not a surjection from~$\set{0,\dots,n-1}$ onto~$\set{0,\dots,n\#n-1}$
(which we will refer to as a surjection from~$n$ onto~$n\#n$).
Our technique uses a new complexity measure that we call the
prefix series for $f(\vec x)$ (Definition~\ref{def:ps}) which might be useful in future work. 
It should be noted that $\ST1$ can prove the surjective pigeonhole
principle for~$n$ onto~$n^2$ from the principle for~$n$ onto~$n\#n$.
However,
the amount of iteration takes one (just barely) out of the class~$A^3$.

We now discuss the organization of the rest of the paper and give a
high-level sketch of the proof. In the next section we introduce the necessary notations from bounded arithmetic 
and define our function algebras. 
In Section~\ref{sec:pfx-series} we define the notion of a ``prefix series.''
Roughly speaking, a prefix series for $f(\vec x)$ is a representation of
$f(\vec x)$ as a difference of sums of prefixes of the values~$\vec x$.
In Theorem~\ref{complexity} we establish a bound on the length
of such prefix series.  In Section~\ref{sec:bit-series} we convert
the prefix series representation to one in which the prefixes are replaced
by bits.  We compute a bound on the length of such a representation
and combine it with Theorem~\ref{complexity} to compute a bound on the
number of times the binary representation of $f(\vec x)$ can alternate
between~$0$ and~$1$ (Lemmas~\ref{convert} and~\ref{numblocks}).  
For $f\in A^3$ this bound is provably lower than the number
of alternations in $\lfloor (n\#n-1)/3\rfloor$, allowing us to conclude
that~$f$ is not a surjection from~$n$ onto~$n\#n$
(Theorem~\ref{clm:php3}).  We conclude with some remarks on generalizations
and extensions.

\section{Preliminaries}
This paper assumes familiarity with the texts of
either Buss~\cite{Buss:Bounded-Arith}, \Krajicek~\cite{krajicek:Bounded-Arithmetic}, or \Hajek\ and \Pudlak~\cite{hajek-pudlak:Metamathematics}. 
For completeness, we review the basic notations of
bounded arithmetic. The specific bootstrapping we are following is that of
Pollett~\cite{pollett:apal99}, but yields equivalent theories to the ones in the books just mentioned. 
The language $L_2$
contains the non-logical symbols $0$, $S$, $+$, $\cdot$, $\mathord=$,
$\leq$, $\monus$, $\HALF{x}$, $\lh x$, $\MSP(x,i)$ and $\#$. The
symbols $0$, $S(x) =x+1$, $+$, $\cdot$, and $\leq$ have the usual
meaning. The intended meaning of $x \monus y$ is $x$ minus $y$ if
this is greater than zero  and zero otherwise, $\HALF{x}$  is $x$
divided by $2$ rounded down, and $\lh x$  is $\lceil \log_2(x +1)
\rceil$, that is, the length of $x$ in binary notation. $\MSP(x,i)$
stands for  `most significant part' and is intended to mean
$\lfloor x/2^i \rfloor$. Finally, $x \# y$ reads `$x$ smash $y$'
and is intended to mean $2^{\lh x\lh y}$. 
The original formulations of bounded arithmetic
do not usually include  $\MSP(x,i)$ and $\monus$, but instead define them with
formulas. One advantage to our approach is that one can define terms in the
language to do a limited amount of sequence coding, which allows us to
more directly formulate our principles in the language~$L_2$.

The bounded formulas of~$L_2$ are classified into hierarchies~$\SIG i$
and~$\PI i$ by counting alternations of quantifiers, ignoring sharply-bounded
quantifiers, analogous to the hierarchies~$\Sigma^0_i$ and~$\Pi^0_i$
of the arithmetic hierarchy.
Here sharply bounded means bounded by a term of the
form $\lh t$.
Formally, a $\SIG 0$ ($\PI 0$) formula is one in which all quantifiers
are sharply-bounded.  The $\SIG{i+1}$ ($\PI{i+1}$) formulas contain
the $\SIG i\cup\PI i$ formulas and are closed under $\NOT A$,
$A\IMP B$, $B\AND C$, $B\OR C$, sharply-bounded quantification, and
bounded existential (universal) quantification, where
$A$ is $\PI {i+1}$ ($\SIG {i+1}$) and $B$ and $C$ are $\SIG {i+1}$ 
($\PI {i+1}$).

The theory $\BASIC$ is axiomatized
by a finite set of quantifier-free
axioms for the non-logical symbols of $L_2$.
$\IND^{\tau}$ consists of formulas of the form
\[
  A(0) \AND (\forall x)(A(x) \IMP A(Sx)) \IMP (\forall x) A(\ell(x)).
\]
for $\ell \in \tau$ where $\tau$ is collection of unary functions. Let $id$ denote the identity
function. 
$\mathcal C$-$\IND$ and -$\LIND$ (\emph{length} induction)
are obtained by taking
$A\in\mathcal C$ and
$\tau$ to be $\{\id\}$ and $\{\lh{\id}\}$, respectively
(we will write $\lh{\id}$ for $x\mapsto \lh{\id(x)}$, etc.).  
The theories $\TT{i}$ and $\ST{i}$
are axiomatized as $\BASIC$ together with respectively $\SIG{i}$-$\IND$
and $\SIG{i}$-$\LIND$.

We next briefly consider sequence coding and bit manipulation in our systems of arithmetic.
The term $\Bit{i,w} := \MSP(w,i)\monus 2\cdot\lfloor \MSP(w, i)/2\rfloor$
is the $i$-th bit of~$w$.  
The ordered pair~$\pair{x}{y}$ can be defined as the binary string
$1\langle x\rangle1\langle y\rangle$ where
$\langle x\rangle$ is the binary representation of~$x$ padded with $0$'s
on the left to have length~$\lh x+\lh y$ and similarly for~$\langle y\rangle$.
Sequences can be defined as ordered 
pairs in which the first component
specifies a block size and the second a concatenation of blocks.
The predicate~$\Seq(s)$ that is true when~$s$ is the code of a sequence
can be given a~$\SIG0$-definition.
The function $\SqBd(a, b):= 64(2\#a\#(2(2b+1)))$ is a bound on the value of any sequence of
length~$<\lh{b}$, 
each of whose components is~$\leq a$, and
$\proj{b}{w}$ is defined to be the $b$-th element of the sequence $w$.
$\proj bw$ can be defined as a term in our language, and the basic properties
of $\SqBd$ and $\proj bw$ can be proved using open length induction. 
We will use sequences of pairs extensively in this paper, 
so define the term $\PSqBd(a, b) = \SqBd(\SqBd(a,2^2),b)$ that is
a bound on the value of any sequence of pairs of length~$<\lh{b}$ for
which each component of each pair is~$\leq a$.

The theory $\ST1$
can prove the existence of sequences and properties of sequences using length induction
if particular elements in the sequence have $\SIG1$-definitions. Sometimes it will be convenient
to use other principles more directly connected to sequences. It is known that $\ST1$ can prove the
following $\SIG1$-$\REPL$ principle (see \cite{Buss:Bounded-Arith} or \cite{pollett:phd}):
\[
\forall x\leq\lh b\exists y\leq a A(x,y) \IMP
\exists w\leq \SqBd(a,2b+1)\forall i\leq \lh b\bigl(
  \proj iw\leq a \AND A(x,\proj iw)
\bigr).
\]
where $A$ is a $\SIG1$-formula. Using this principle, we can $\SIG1$-define the sequence $\langle f(0,x), f(1,x), \ldots, f(p(\lh x), x)\rangle$ where $p$ is a polynomial provided we know $f(i,x)$ is $\SIG1$-definable (see below). 
Further it can be shown that $\ST1$ can prove basic properties of this sequence. The $\SIG1$-$\REPL$ scheme can be used to prove another useful scheme in
$\ST1$, that of  $\SIG1$-$\COMP$
\begin{equation*}
  (\exists w < 2^{{\lh a}})(\forall i < {\lh a})(A(i,a) \IFF \BIT(i,w)=1).
\end{equation*}
which allows one to get a bit-string of values for a $\SIG1$-formula $A(i,a)$.

The $\IND^{\tau}$ scheme is 
closely connected with the following type of bounded primitive recursion:
\begin{defn}
\label{defn:bpr}
($\BPR^{\tau}$) Let $\tau$ be a set of unary functions.
$f$ is defined from functions $g$, $h$, $t$ and $r$ by 
\emph{$\tau$-length bounded primitive recursion} if:
\begin{align*}
F(0,\vec{x}) &= g(\vec{x})\\
F(n+1, \vec{x}) &= \min(h(n,\vec{x}, F(n,\vec{x})), r(n,\vec{x}))\\
f(n,\vec{x}) &= F(\ell(t(n, \vec{x})),\vec{x})
\end{align*}
for some $r,t \in L_2$ and $\ell \in \tau$.
\end{defn}

Let $L^-_2$ be the language of $L_2$ where the symbol for multiplication has been replaced with $\PAD(x,y)$ with intended meaning
$x\cdot 2^{\lh y}$. As $\PAD$ is definable with an $L_2$-term any $L^-_2$-term can be rewritten as an $L_2$-term. Given a class of formulas $\Psi$, we say an arithmetic theory $T$ can {\em $\Psi$-define} a function
$f$ if there is a formula $A_f$ in $\Psi$ such that $T$ proves:
\begin{enumerate}
\item $T\proves \forall x \exists! y A_f(x,y)$
\item $\nat \models \forall  x A_f(x, f(x))$
\end{enumerate}

\begin{defn}
For a set~$\tau$ of function symbols, the set $A^\tau$ is defined as follows:
\begin{enumerate}
\item The function symbols of~$L_2^-$ are in~$A^\tau$ along with
symbols~$\pi^n_i$ for $0\leq i<n$ (intended interpretation:  projections);
\item If $f,g_1,\dots,g_r\in A^\tau$ and $f$ is $r$-ary, then
$C_{f,g_1,\dots,g_r}\in A^\tau$ (intended interpretation:  the
composition of $f$ with $g_1,\dots,g_r$);
\item If $g,h\in A^\tau$, $t,r\in L_2^-$, and $\ell\in\tau$, then
$R_{g,h,t,r,\ell}\in A^\tau$ (intended interpretation:  the function
defined by $\ell$-bounded primitive recursion from~$g$, $h$, $t$,
and $r$).
\end{enumerate}
\end{defn}

$A^\tau$ of course corresponds to a function algebra and we shall
frequently informally refer to it as such.
We write $A^m$ for $A^{\{\lh{\id}_m\}}$; we shall focus primarily
on these classes in all but the last section.
Pollett~\cite{pollett:phd} considers these classes where the initial functions also include multiplication. 
In particular, it is known that $A^1$ corresponds to the polynomial time functions
and Pollett shows that $A^4\subset A^1$ as $A^4$
cannot define $\lfloor x/3\rfloor$.  
When we refer to terms (formulas, etc.) over~$A^\tau$ in, e.g., $\ST1$,
we assume that the functions in~$\tau$ are defined by~$L_2^-$ terms
and that the defining axioms of the functions symbols are
(conservatively) added to the theory (we shall always
have $A^\tau\subseteq A^1$).
Using the close connection between $\LIND$ and $\BPR^{\{\lh{\id}\}}$ 
Buss~\cite{Buss:Bounded-Arith} shows that the functions
in $A^1$ are precisely the functions $\SIG1$-defined in $\ST1$.

A couple of notations that we use frequently in this paper are:
\begin{itemize}
\item For $\vec x=x_1,\dots,x_k$, $\vec x<n$ abbreviates
  $x_1<n\AND\dots\AND x_k<n$.
\item We will write $\#^b(n)$ for $n\#\dots\# n$ ($b-1$ $\#$'s).
\end{itemize}

%%%%%
%%%%% PHP principles
%%%%%
\begin{defn} ~
\label{defn:php}
\begin{enumerate}
\item For a unary function symbol~$f$, $s\PHP(f)^m_n$ is the formula
$n<m \AND \exists y< m\forall x< n f(x)\not= y$.  
\item The
\emph{weak surjective pigeonhole principle for~$f$}, $s\WPHP(f)$, is the
sentence~$\forall n.s\PHP(f)^{n^2}_n$.  If~$A$ is a set of function
symbols, $s\WPHP(A)$ is the set of formulas $s\WPHP(f)$ for 
unary functions $f\in A$.
\item 
The sentence $s\WPHP^\#(f)$ is
$\forall n.s\WPHP^{n\# n}_n(f)$ and $s\WPHP^\#(A)$ is defined similarly.
\end{enumerate}
\end{defn}

\begin{prop}
If $A$ is a set of function (symbols) closed under
$\BPR^{\set{\lhlh{\id}}}$, then
$\ST1\proves s\WPHP^\#(A)\IMP s\WPHP(A)$.
\end{prop}
\begin{proof}
If $f_0$ is a surjection from~$2^{\lh m}$ onto $2^{2\lh m}$, 
define surjections $f_r$ from~$2^{\lh m}$ onto~$2^{2^{r}\lh m}$
by setting $f_{r+1}(x)$ to be the result of replacing each length-$\lh m$
block~$y$ of $f_r(x)$ by~$f_0(y)$.
Then $f_{\lhlh m}$ is a surjection from~$2^{\lh m}$ onto~$2^{\lh m\lh m}$.
\end{proof}

\section{Prefix series representation}
\label{sec:pfx-series}
In this section we introduce the notion of a prefix series, 
which is our main technical tool for proving the weak surjective pigeonhole
principle.

%%%%%
%%%%% Prefix series, bit series, summand complexity
%%%%%
\begin{defn}
\label{def:ps}
~
\begin{enumerate}

\item \label{item:ps}
A \emph{prefix series for~$M$ from~$\vec m$ of width~$w$ and length~$k$} is
a pair $\pair PN$ of sequences such that:
\begin{enumerate}

\item $P=\langle\pair{a_0}{b_0},\dots,\pair{a_{k_P-1}}{b_{k_P-1}}\rangle$
and $N=\langle\pair{c_0}{d_0},\dots,\pair{c_{k_N-1}}{d_{k_N-1}}\rangle$;

\item $M = \sum_{i=0}^{k_P-1}a_i2^{b_i}\monus
\sum_{i=0}^{k_N-1}c_i2^{d_i}$;

\item For all~$i$, $b_i, d_i\leq \lh{w}$;

\item $k = k_P+k_N$;

\item For all~$i$, either $a_i=1$ or there are~$j$ and~$y\leq\lh{m_j}$
such that $a_i=\MSP(m_j,y)$ and similarly for~$c_i$.
\end{enumerate}

\item 
A \emph{bit series for~$M$ from~$\vec m$ of width~$w$ and length~$k$} is a
prefix series for~$M$ from~$\vec m$ of width~$w$ and length~$k$ in which 
all~$a_i$'s and $c_i$'s are~$1$.

\item For terms~$t(\vec x)$ and~$w(n)$
let $k_{t,w}(n)$ be the least~$k$ such that if $m_i<n$ for all~$i$, then
there is a prefix series for~$t(\vec m)$ from~$\vec m$ of 
width~$\leq w(n)$ and length~$\leq k$.
Then $k_{t,w}$ is the \emph{$w$-summand complexity} of~$t$
($k_{t,w}(n)$ may not be defined for all~$w$). 
\end{enumerate}
\end{defn}

For any function~$f$, if we could define the
term~$w(n) = \max\setst{\lh {f(\vec x)}}{\vec x<n}$, then
$w(n)$ itself would be a bound on~$k_{f(\vec x),w}$:  just use
the binary representation of~$f(\vec x)$ to define a bit-series.  Of course,
such a term~$w$ is problematic;
our first goal is to show that for every~$f\in A^3$ there is in fact
a~$w$ such that
$k_{f(\vec x),w}$ has a ``tractable'' upper bound (and in particular
is defined).

%%%%%
%%%%% PS, PSBound, BS, BSBound
%%%%%
\begin{defn} ~
\begin{enumerate}
\item 
$\BPS(S, y, x_1,\dots,x_r, w, k, \delta)$ is the predicate
\begin{multline*}
S<\PSqBd(x_1+\dots+x_p+\lh{w},2^{\min(k,\lh\delta)})\AND
\forall i<\min(k,\lh\delta)\Bigl[ \\
\exists a,b<\proj iS\Bigl(
\proj iS=\pair a b \AND
\Bigl(a=1\OR\bigvee_{j=1}^r \bigl(\exists r<\lh{x_j}(a=\MSP(x_j,r))\bigr)\Bigr)\AND b<\lh{w}\AND \\
\eval(S) = y\Bigr)
\Bigr]
\end{multline*}
that states that $S$ is a prefix series for~$y$ from~$\vec x$ of
width~$w$ and length~$\min(k,\lh\delta)$.
Here $\eval$ is the polynomial-time function
that on input~$\pair PN$ as in Definition~\ref{def:ps}(\ref{item:ps})
outputs $\sum_{i=0}^{k_P-1}a_i2^{b_i}\monus
\sum_{i=0}^{k_N-1}c_i2^{d_i}$.  Note that
$\exists S.\BPS(S,y,\vec x,w,k,\delta)$ is a $\SIG1$~formula.
We discuss the parameter~$\delta$ below.

\item 
Let $t(\vec x)$, $w(n)$, $k(n)$, and $\delta(n)$
be terms.  $\PSBOUND{t}w{k,\delta}$ is the predicate
\[
  \exists n_0\forall n\geq n_0\forall\vec x<n\exists S.\BPS(S, t(\vec x), \vec x, w(n), k(n),\delta(n))
\]
that states that for sufficiently large~$n$,
$\min(k(n),\lh{\delta(n)})$ is an upper bound on~$k_{t,w}(n)$ (and in particular, $k_{t,w}(n)$
is defined).

\item \label{item:bs}
$\BS(S, y, w, k,\delta)$ is the predicate
\[
S<\PSqBd(1+\lh{w},2^{\min(k,\lh\delta)})\Bigl[
\forall i<\min(k,\lh\delta)\exists b<\lh w\Bigl(
\proj iS=\pair 1 b \AND
\eval(S) = y\Bigr)
\Bigr]
\]
that states that $S$ is a bit series for~$y$ of width~$w$ and
length~$\min(k,\lh\delta)$.  $\BSBOUND tw{k,\delta}$ is defined analogously to
$\PSBOUND ts{k,\delta}$.
\end{enumerate}
\end{defn}

The point behind the parameter~$\delta$ is to ensure that the exponentiation
terms in~$\BPS$ and $\BS$ are bounded by~$L_2$-terms.
Our goal is now the following:  given an $A^3$-function
symbol~$f$, find $L_2$-terms $w$, $k$, and~$\delta$
such that $\ST1\proves \PSBOUND {f\vec x}w{k,\delta}$; in
other words, find a bound on the lengths of the prefix series for~$f(\vec x)$.
In fact, the form of~$k$ will be made explicit, and this will allow
us to take $\delta=n^2$ for all function symbols in~$A^3$.
However, for some preliminary observations which do not rely on the
form of~$k$, we must allow this parameter to vary.

\begin{lem}
\label{clm:longer-width-length}
$\ST1$ proves the following:
\[
\BPS(S,y,\vec x,w,k,\delta)\AND
w\leq w'\AND k\leq k'\IMP
\BPS(S,y,\vec x,w',k',\delta
).
\]
In particular, for any terms~$t$, $w$, $w'$, $k$, $k'$, and~$\delta$,
\[
\ST1\proves(\exists n_0\forall n\geq n_0(w(n)\leq w'(n) \AND k(n)\leq k'(n))\AND\PSBOUND tw{k,\delta})\IMP
\PSBOUND t{w'}{k',\delta}
\]
and similarly for the bit-series predicates.
\end{lem}

\begin{lem}
\label{clm:L2-lh-bd}
For every $f(\vec x)\in A^1$ there is an
$L_2$-term~$s(\vec x)$ without $\monus$ or $\MSP$ (hence monotone) such that 
$\ST1\proves \forall\vec x (f(\vec x)\leq s(\vec x))$.  In particular,
there is a number~$b$ such that
$\ST1\proves\exists n_0\forall n\geq n_0\forall\vec x<n(\lh{f(\vec x)}\leq\lh n^b)$.
\end{lem}
\begin{proof}
The first part is proved by induction on the definition of~$f$.  The
base cases are immediate (bound $x\monus y$ and $\MSP(x,y)$ by~$x$)
and composition is handled by substitution.
Suppose $f$ is defined as in Definition~\ref{defn:bpr}; the induction
hypothesis gives us bounds~$u_g$, $u_t$, and $u_r$ for $g$, $t$, 
and~$r$ respectively.
Then $F(y,\vec x)\leq u_g(\vec x)+u_r(y-1,\vec x)$ and hence
$f(x,\vec x)\leq u_g(\vec x)+u_r(\lh{u_t(x,\vec x)},\vec x)$.
For the second part,
prove that for any~$L_2$-term~$u(\vec x)$ without~$\monus$ or~$\MSP$
there is a number~$b$
such that $\lh{u(\vec x)}\leq\lh n^b$ for sufficiently large~$n$ and
$\vec x<n$ by induction on~$u$.  For example, if $u = u_1\#u_2$,
then take $b=b_1+b_2$, where $b_i$ is the inductively-given exponent
for~$u_i$.
\end{proof}

\begin{lem}
\label{clm:bps-bound}
$\ST1\proves\forall x\exists S.\BS(S,x,x,\lh{x},x)$.
In particular, for every~$A^1$-term~$u(\vec x)$ there is a number~$b$ such that
$\ST1\proves \BSBOUND{u}{\lh{n}^b}{\lh n^b,\#^b(n)}$.
\end{lem}
\begin{proof}
For the first part use~$\SIG0$-$\REPL$ to construct the sequence
of pairs~$\pair {\BIT(i,x)}{i}$ such that $\BIT(i,x)=1$, 
which witnesses the claim.  
For the second part, fix any~$\vec x$; then there is an~$S$
such that $\BS(S, u(\vec x), u(\vec x), \lh{u(\vec x)},u(\vec x))$.  Now
take $n_0$ and~$b$ as in Lemma~\ref{clm:L2-lh-bd} and apply
Lemma~\ref{clm:longer-width-length}.
\end{proof}

We call the bit series given in Lemma~\ref{clm:bps-bound} the
\emph{natural} bit series for~$x$.
We need the following bound for calculating the length of a prefix
series for (the function represented by) an~$\MSP$-term:

\begin{lem} 
\label{clm:msp-error}
The following are provable in~$\ST1$:  for any~$\vec a$, any
length~$k$ and any length~$y$:
\begin{enumerate}
\item \label{item:half}
$\sum_{i=0}^{k-1}\MSP(a_i,1) \leq \MSP(\sum_{i=0}^{k-1}a_i,1)
\leq \left(\sum_{i=0}^{k-1}\MSP(a_i,1)\right)+k-1$.
\item \label{item:general}
$\sum_{i=0}^{k-1}\MSP(a_i,y) \leq \MSP(\sum_{i=0}^{k-1}a_i,y)
\leq \sum_{i=0}^{k-1}\MSP(a_i,y)+\sum_{i=0}^{y-1}\MSP(k,i)$.
\item \label{item:monus}
$\MSP(a,y)\monus\MSP(b,y)\monus 1\leq \MSP(a\monus b,y)\leq
\MSP(a,y)\monus \MSP(b,y)$.
\end{enumerate}
\end{lem}

\begin{prop}
\label{clm:pfx-series-sum}
$\ST1$ proves the following:
\begin{multline*}
\exists S''\bigl[
\bigl(\PS(S,y,\vec x,w,k,\delta)\AND
\PS(S',y',\vec x,w',k',\delta')\bigr)\IMP \\
\PS(S'',y+y',\vec x,w+w',k+k',\delta\delta')\bigr].
\end{multline*}
The same claim holds with~$y+y'$ replaced with~$y\monus y'$.
\end{prop}
\begin{proof}
Working in~$\ST1$,
suppose $y = P\monus N$, and
$y' = P'\monus N'$ are prefix series for~$y$ and~$y'$ of
widths~$w$ and~$w'$ and lengths~$k$ and~$k'$ respectively.
If $N\geq P$, then
$y+y' = P'\monus N$.
If $N<P$ and $N'\geq P'$, then $y+y' = P\monus N$.  If $N<P$ and
$N'<P'$, then $y+y'= (P+P')\monus(N+N')$.  
In each case, the width and length of the
prefix series are at most $w+w'$ and $k+k'$ respectively.
\end{proof}

We shall frequently rearrange sums of differences of sums
in this way to obtain prefix series; we will not frequently point
out that we are doing so.

\begin{prop}
\label{clm:pfx-series-msp}
$\ST1$ proves the following:
\begin{multline*}
\exists S'\bigl[
  \PS(S,z,\vec x,w,k,\delta)\IMP \\
  \PS(S',\MSP(z,y),\vec x,w+\lh k+\lhlh k,k+\lh k+\lhlh k,\delta\lh\delta\lhlh\delta)
\bigr].
\end{multline*}
\end{prop}
\begin{proof}
Suppose $P\monus N$ is a prefix series for~$z$ from~$\vec x$
of width~$w$ and length~$k$ as in Definition~\ref{def:ps}(\ref{item:ps}).
From Lemma~\ref{clm:msp-error} and arithmetic we have that
$Q \monus k\lh k \leq \MSP(P\monus N,y) \leq Q+k\lh k$
where 
$Q = \sum_{i=0}^{k_P-1}\MSP(a_i2^{b_i},y)\monus\sum_{i=0}^{k_N-1}\MSP(c_i2^{d_i})$.
Thus there is some~$e\leq k\lh k$ such that
$\MSP(z, y) = Q\monus e$ or $\MSP(z,y) = Q+e$.  Since~$Q$ is a prefix series
from~$\vec x$ of width~$\leq w$ and length~$\leq k$,
by Proposition~\ref{clm:pfx-series-sum} and Lemma~\ref{clm:bps-bound}
there is a prefix series
for~$\MSP(z,y)$ from~$\vec x$ of width~$w+\lh k+\lhlh k$ and
length~$k+\lh k+\lhlh k$.
\end{proof}

We now set about showing that for $m\geq 3$ and every function symbol
$f\in A^m$ there is an~$L_2$-term~$w_f(n)$ and
a number~$b_f$ such that if $k(n)$ is the term~$\lhlh{n}^{{\lh n}_m^{b_f}}$
then $\ST1\proves\PSBOUND{f\vec x}{w(n)}{k(n),n^2}$.  
More precisely, we will write $\lhlh{n}^{\lh n_m^b}$ for the
term $\lhlh n\# \left(\#^b(\lh n_{m-1})\right)$
so that $k(n)$ is an $L_2$-term.  It is also
easy to see that if $m\geq 3$, then $\ST1$ proves
that $\lhlh n^{\lh n_m^b}$ is bounded
by $2^{\lh n_3^{b+1}}$, which in turn is bounded above
by $\lh{n^2}$ for sufficiently large~$n$ (where the point at which this
holds depends only on~$b$).  Thus from now on, we shall simply
write $\PSBOUND{f\vec x}{w(n)}{k(n)}$ with the bounding term
always implicitly~$n^2$.
The proof is by induction on the definition of~$f$; we separate out the
base case into its own proposition.

\begin{prop}
\label{functionbounds}
If $f$ is an $L_2^-$-function symbol, 
then there is an $L_2$-term~$w$ and a number~$b$ such that
$\ST1\proves \PSBOUND{f\vec x}{w(n)}{\lhlh n^b}$.
\end{prop}
\begin{proof}
The proof is a straightforward analysis; most cases are handled
by already-proved lemmas and propositions.
If $f=0$, then $w_f = k_f = 0$ and if
$f=x\# y$ then we can take $w(n) = n\#n$ and $k=1$ since
$fxy = 1\cdot2^{\lh x\lh y}$.
If $fx = \lh{x}$ then an argument as in Lemma~\ref{clm:bps-bound} applies
using Lemma~\ref{clm:L2-lh-bd} to bound $f(x)$ by $\lhlh n^b$.
If $f(x,y) = x+y$ or $f(x,y) = x\monus y$ then
Proposition~\ref{clm:pfx-series-sum} applies and
if $f(x,y) = \MSP(x, y)$ then Proposition~\ref{clm:pfx-series-msp} does.
If $f(x,y) = \PAD(x, y)$, then a prefix series for $f(x,y)$ from~$x,y$
is given by $x\cdot2^{\lh y}$, which has width~$\leq \lh n$ and length~$1$.
\end{proof}

\begin{thm}
\label{complexity}
If $m\geq 3$ and $f$ is an~$A^m$-function symbol then there is an
$L_2$-term~$w_f$ and a number~$b_f$ such that
$\ST1\proves\PSBOUND{f\vec x}{w(\vec x)}{k(n)}$, where
$k(n) = \lhlh{n}^{\lh{n}_m^{b_f}}$.
\end{thm} 
\begin{proof}
The proof is by induction on the definition of~$f$.  The base case
in which $f$ is an $L_2^-$ symbol is handled
in Proposition~\ref{functionbounds}.

Suppose $f$ has defining equation $f\vec x = g(h_1\vec x,\dots,h_r\vec x)$.
By the induction hypothesis we have terms~$w_h(n)$, $k_h(n)$,
$w_g(n)$, and $k_g(n)$ such that
$\ST1\proves\bigwedge_i\PSBOUND{h_i\vec x}{w_h}{k_h}\AND
\PSBOUND{g\vec x}{w_g}{k_g}$.  Let $n_h$ be such that for all $n\geq n_h$
there is a prefix series for $h_i(\vec x)$ from~$\vec x<n$ of the given
width and length, and define $n_g$ similarly.
Furthermore take a constant~$B$
such that if $n\geq n_h$ and $\vec x<n$, then $\lh{h\vec x}\leq\lh n^B$.
Take $n_0$ large (we shall impose constraints as the proof progresses),
$n\geq n_0$, and $\vec x<n$.  The induction hypothesis for~$h_i$ gives
us a prefix series~$S_i$ for $h_i(\vec x)$ from~$\vec x$ of width~$w_h(n)$
and length~$k_h(n)$ (assume $n_0\geq n_h$).  Since $n_0\geq n_h$ we
also have that $h_i(\vec x)\leq 2^{{\lh n}^B}$.
Now the
induction hypothesis for~$g$ gives us a prefix series~$S_g$ for
$g(h_1(\vec x),\dots,h_r(\vec x))$ from~$h_1(\vec x),\dots,h_r(\vec x)$
of width~$w_g(2^{\lh n^B})$ and length~$k_g(2^{\lh n^B})$
(assume $2^{\lh{n_0}^B}\geq n_g$).  The terms in~$S_g$ have the
form~$\MSP(h_i(\vec x),y)2^j$ for some~~$i$, $y$ and~$j$ (the terms
with coefficient~$1$ we leave as they are).  Replace each such term
with a prefix series for~$\PAD(\MSP(h_i(\vec x),y),2^{j-1})$
from~$\vec x$; this is obtained from the inductively-given
prefix series by Lemma~\ref{clm:pfx-series-msp} and then padding,
and has width at most~$w_h(n)+w_g(2^{{\lh n}^B})$ and
length at most~$k_h(n)+\lh{k_h(n)}+\lhlh{k_h(n)}$.  After replacing
all terms in~$S_g$ in this way and rearranging if necessary (dropping
expressions that evaluate to~$0$) we obtain a prefix series~$S$
for $g(h_1(\vec x),\dots,h_r(\vec x))$ from~$\vec x$ of width
at most~$w_h(n)+w_g(2^{\lh n^B})$ and length
$k_g(2^{\lh n^B})(k_h(n)+\lh{k_h(n)}+\lhlh{k_h(n)})$.  Finally,
by taking~$n_0$ large enough, $k_g(2^{\lh n^B})$ is
bounded above by $\lhlh{n}^{\lh{n}_m^{b_g+1}}$, from which an
upper bound on the length of the correct form is easily obtained,
completing the proof for this case.

Suppose $f$ is defined by $\lh{\cdot}_m$-bounded recursion from~$g$,
$h$, $t$, and~$r$ with intermediate function~$F$ as in
Definition~\ref{defn:bpr}.
Let $k'(n) = k_h(n) + \lhlh n^{b_r}$.
Take~$b$ and~$c$ such that for sufficiently large~$n$ and~$y,\vec x<n$,
$\lh{F(y,\vec x)}\leq\lh n^b$ and $\lh{t(x,\vec x)}_m\leq \lh n_m^c$.  
Now take a sufficiently large~$n_0$,
$n\geq n_0$, $\vec x<n$, and show by length-induction on~$y<n$
that there is a prefix series for~$F(y,\vec x)$ from~$y,\vec x$
of length~$(3k'(2^{\lh n^b}))^yk_g(n)$.  For the induction step, since
$F(y+1,\vec x)$ is
defined as a composition of $h$ with $F(y,\vec x)$ (the case in
which $F(y+1,\vec x) = r(y,\vec x)$ is immediate) an
argument as in the previous case applies.  
Now taking $y=\lh{t(x,\vec x)}_m$ we obtain a prefix
series of length~$(3k'(2^{\lh n^b}))^{\lh{n}_m^c}k_g(n)$ which
we can bound by a term of the form~$\lhlh n^{\lh n_m^{Bc+b_g+1}}$
where $B = b_h+1$.  Similarly we obtain a bound on the width
of the prefix series for $F(y,\vec x)$ of the form
$yw'(2^{\lh n^b})w_g(n)$ where $w'(n) = w_h(n) + w_r(n)$; 
when $y=\lh{t(x,\vec x)}_m\leq\lh n_m^c$, we obtain an
term bounded by an $L_2$-term in~$n$.
\end{proof}

\section{Bit series representation and the weak pigeonhole principle}
\label{sec:bit-series}
We now extract bounds on lengths of bit series representations from
bounds on prefix series representations and use them to determine bounds
on the number of times the binary representation of~$f(\vec x)$ can
alternate between~$0$'s and $1$'s.

\begin{lem}
\label{convert}
For any terms~$t$, $w$, $k$, and~$\delta$,
$\ST1\proves\PSBOUND tw{k,\delta}\IMP \BSBOUND t {w(n)+\lh n} {\lh nk(n),n\#\delta(n)}$.
\end{lem}
\begin{proof}
Given a prefix series for~$t(\vec x)$ from~$\vec x$, 
replace each term~$a2^b$ with
$1\cdot2^{b+i_1}+\dots+1\cdot2^{b+i_r}$, where the $i_1,\dots,i_r$ are
exactly those~$i$ such that $\BIT(i,a)=1$.  
Since each $i_j\leq\lh a\leq\lh x_l\leq\lh n$ for some~$l$, the resulting
bit series has width at most $w(n)+\lh n$.
Since $\lh a\leq\lh n$
each term is replaced with a summand of at most $\lh n$ terms.  Since there
are at most $k(n)$ summands, the resulting bit series for~$t(\vec x)$
from~$\vec x$ has length at most $\lh nk(n)$.
\end{proof}

Given the binary expansion of a number $y$, a {\em block} is a substring of all 0's or all 1's of maximal length.  Let $\#_B(y)$ denote the number of blocks in $y$'s binary expansion. This number can be $\SIG1$-defined in $\ST1$ as 
$(\#i\leq \lh y)(\BIT(i,y) \neq \BIT(i+1,y))$.
Here $(\#i \leq\lh y)B$ is the operator which counts the number of $i\leq \lh y$ such that $B$ holds. It is known to be $\SIG1$-definable in $\ST1$ provided $B$ is $\DELT1$ by Buss~\cite{Buss:Bounded-Arith}.

\begin{lem}
\label{numblocks}
$\ST1$ proves the following:
\begin{multline*}
\forall w\delta\forall k\forall S<\PSqBd(1+\lh w,2^{\min(\lh k,\lh\delta)})\bigl[\\
\BS'(S,\eval(S), w,\lh k,\delta)\IMP(\#_B(\eval(S)) \leq 2\lh k+1)
\bigr]
\end{multline*}
where $\BS'$ is the part of the definition of~$\BS$ 
(Definition~\ref{def:ps}(\ref{item:bs})) in brackets.  In other words,
the binary expansion of
a number represented by a bit-series of length~$\lh k$ has at
most~$2\lh k+1$ blocks.
\end{lem}
\begin{proof}
Fix~$w$ and~$\delta$; we prove this $\PI1$ claim by length-induction
on~$k$.
If $k=0$ then $\eval(S)=0$ and the claim is immediate, so assume the claim is
true for~$k$ and that $\BS'(S,w,k+1,\delta)$.  Then $\eval(S)=\eval(S')\pm 2^j$ 
for some~$S'$ satisfying $S'<\PSqBd(1+\lh w,2^{\min(k,\lh\delta)})$
and $\BS'(S',\eval(S'),w,k,\delta)$, 
so the induction hypothesis applies to~$S'$.
It is now a matter of exhausting cases on whether $\eval(S)=\eval(S')+2^j$
or $\eval(S)=\eval(S')\monus 2^j$ and 
$\BIT(j-1,\eval(S'))$, $\BIT(j,\eval(S'))$, and $\BIT(j+1,\eval(S'))$
to show that $\#_B(\eval(S))\leq \#_B(\eval(S'))+2$, from which the claim follows.
\end{proof}

\begin{thm}
\label{clm:php3-prelim}
For any $f\in A^3$, $\ST1\proves\exists n_0\forall n\geq n_0.s\PHP^{n\#n}_n(f)$.
\end{thm}
\begin{proof}
Combining Theorem~\ref{complexity} with Lemmas~\ref{convert}
and~\ref{numblocks} we have that for sufficiently large~$n$,
if $\vec x<n$ then $\#_B(f(\vec x))\leq 4\lh n\lhlh n^{\lh n_3^b}$ for
some fixed number~$b$.  Now $\ST1$ proves
that $\lh n_3^b\leq\floor{\lhlh n/2}$ for sufficiently large~$n$ and
that $\floor{\lh a/2}\leq\lh{\MSP(a,\floor{\lh a/2})}$ for any~$a$; combining these,
we have that 
\[
  \lhlh n^{\lh n_3^b} \leq 2^{\lh n_3^{b+1}} \leq
  2^{\floor{\lhlh n/2}}\leq 2^{\lh{\MSP(\lh n,\floor{\lhlh n/2})}}\leq
  2\MSP(\lh n,\floor{\lhlh n/2}).
\]
Thus we conclude that
$\#_B(f(\vec x))\leq 8\lh n\MSP(\lh n,\floor{\lhlh n/2})$.  Thus
$\lh{\#_B(f(\vec x))}\leq 3+\lhlh n+\floor{\lhlh n/2}\leq
3+\floor{3\lhlh n/2}$.
On the other hand, $\ST1$ proves that if
$a=\lfloor\frac{(n\#n)-1}{3}\rfloor$ then $\#_B(a)\geq \MSP(\lh n^2-1,3)$
(first show that $n\#n-1$ is a
string of all $1$'s, then analyze the grade-school algorithm for
division to show that $\lfloor\frac{n\#n-1}{3}\rfloor=101010\ldots$;
this can be done with open length-induction).
Thus
$\lh{\#_B(a)}\geq \lh{\lh n^2-1}-3\geq 2\lhlh n-3$.
If $\lhlh n\geq 3$ then
$\lh{\#_B(f(\vec x))}\leq \floor{3\lhlh n/2}+3< 2\lhlh n-3 \leq \lh{\#_B(a)}$,
so we conclude that $a\not= f(\vec x)$.
\end{proof}

Finally, we note that the value $n_0$ in Theorem~\ref{clm:php3-prelim} can
be calculated explicitly.  That is, in each argument of this and
the previous section in which the conclusion is of the
form $\ST1\proves\exists n_0\forall n\geq n_0\ldots$, we could have
instead computed a closed term~$N$ and shown
$\ST1\proves\forall n>N\ldots$ (adding~$N$ into the formalism
would have entailed making our already-unpleasant notation even
worse).  Thus we can improve Theorem~\ref{clm:php3-prelim} as follows:

\begin{cor}
\label{clm:php3}
$\ST1\proves s\PHP^\#(A^3)$.
\end{cor}
\begin{proof}
Fix $f\in A^3$.  As just discussed, there is a closed term~$N$ such that
$\ST1\proves \forall n\geq N.s\PHP^{n\#n}_n(f)$.  Since $N$ is a closed term,
for each~$M<N$ there is an explicit proof in~$\ST1$ of
$s\PHP^{M\#M}_M(f)$, and hence we conclude that 
$\ST1\proves s\PHP^{\#}(f)$.
\end{proof}

\section{Generalizations and extensions}
\label{sec:extensions}

Analyzing the details of the above proofs, we can determine the
properties of~$\lh{\id}_3$ that are required in order to generalize
the result to function classes~$\tau$.  The key point
is that $(\lh n_3)^{b}\in o(\lhlh n)$:

\begin{thm}
\label{canprove}
Let~$\tau$ consist of unary functions~$\ell$ such that:
\begin{enumerate}
\item For every~$\ell\in\tau$ there is a constant~$N$ such that 
$\ST1\proves\forall n\geq N(\lh n_3\ell(n)\leq\floor{\lhlh n/2})$.
\item For every~$\ell_1,\ell_2\in\tau$, there is~$\ell_3\in\tau$ 
and a number~$N$ such
that $\ST1\proves\forall n\geq N(\ell_1(n)+\ell_2(n)\leq \ell_3(n))$.
\item For every~$\ell_1,\ell_2\in\tau$, there is~$\ell_3\in\tau$ 
and a number~$N$ such
that $\ST1\proves\forall n\geq N(\ell_1(n)\ell_2(n)\leq \ell_3(n))$.
\end{enumerate}
Then $\ST1\proves s\PHP^\#(A^\tau)$.
\end{thm}
\begin{proof}
The proofs estimating the lengths of the prefix series carry through 
\emph{mutatis mutandis}, with the new bound on the length being
$\lhlh n^{\ell(n)}$ for some~$\ell\in\tau$; the second two hypotheses
are used in the composition and $\tau$-bounded recursion cases
of Theorem~\ref{complexity}.  The proof of
Theorem~\ref{clm:php3-prelim} relies on the fact that
$\lhlh n^{\lh n_3^b}\leq 2^{\floor{\lhlh n/2}}$.  Now
we need $\lhlh n^{\ell(n)}\leq 2^{\lh n_3\ell(n)}\leq
2^{\floor{\lhlh n/2}}$, which is the first hypothesis.
\end{proof}

Of course, we can add any functions to the algebra~$A^\tau$ provided
that the conclusion of Theorem~\ref{complexity} still holds.  In
particular, if $\ST1$ proves that for sufficiently large~$n$
and $\vec x<n$, $g(x)\leq 2^{\lhlh n^{\ell(n)}}$ then the natural
bit series for $g(\vec x)$ satisfies the conclusion, so any such
functions can be added to~$A^\tau$; we leave it to the reader to 
precisely formulate the corresponding theorem.

Clote~\cite{clote:handbook} gives several interesting function-algebra
characterizations of various complexity classes.  
Most of these
rely on so-called \emph{concatenation recursion on notation} and one other
recursion scheme.  The function~$f$ is defined from~$g$, $h_0$, and~$h_1$
by concatenation recursion on notation if
\begin{align*}
f(0, \vec{x}) &= g(\vec{x})\\
f(2n, \vec{x}) &= s_{h_0(n,\vec{x})}(f(n,\vec{x})), \mbox{ provided $n\neq 0$}\\
f(2n+1, \vec{x}) &= s_{h_1(n,\vec{x})}(f(n,\vec{x}))
\end{align*}
Clote then shows that, for example, the
log-space functions are exactly the closure of~$L_2^-$ under
composition, concatenation recursion on notation, and
sharply-bounded recursion on notation
(called doubly-bounded recursion on notation
by~Clote and Takeuti~\cite{clote-takeuti:FMII}).  This latter scheme defines
a function~$f$ in terms of given functions~$g$, $h_0$, $h_1$, and~$b$ by
\begin{align*}
f(0, \vec{x}) &= g(\vec{x})\\
f(2n, \vec{x}) &= h_0(n,\vec{x}, f(n,\vec{x})), \mbox{ provided $n\neq 0$}\\
f(2n+1, \vec{x}) &= h_1(n,\vec{x}, f(n,\vec{x})) \\
f(n,\vec x) &\leq \lh{b(n,\vec x)}
\end{align*}
It is easy to see that the scheme of weak bounded recursion on
notation preserves the property that for sufficiently large~$n$
and~$\vec x<n$, $f(\vec x)\leq 2^{\lhlh n^{\ell(n)}}$.  Thus, if the
techniques of this paper could be extended to handle concatenation
recursion on notation (for which $\lh{f(\vec x)}$ may now grow
linearly in~$\lh n$), one could hope to prove some version of the
weak pigeonhole principle for these small complexity classes.

\bibliographystyle{abbrv}
\bibliography{master}

\begin{thebibliography}{10}

\bibitem{Buss:Bounded-Arith}
S.~R. Buss.
\newblock {\em Bounded Arithmetic}.
\newblock Bibliopolis, Naples, 1986.

\bibitem{clote:handbook}
P.~Clote.
\newblock Computation models and function algebras.
\newblock In {\em Handbook of computability theory}, volume 140 of {\em Stud.
  Logic Found. Math.}, pages 589--681. North-Holland, Amsterdam, 1999.

\bibitem{clote-takeuti:FMII}
P.~Clote and G.~Takeuti.
\newblock First order bounded arithmetic and small {B}oolean circuit complexity
  classes.
\newblock In {\em Feasible {M}athematics {II} (Ithaca, NY, 1992)}, volume~13 of
  {\em Progr. Comput. Sci. Appl. Logic}, pages 154--218. Birkh\"auser Boston,
  Boston, MA, 1995.

\bibitem{hajek-pudlak:Metamathematics}
P.~H{\'a}jek and P.~Pudl{\'a}k.
\newblock {\em Metamathematics of First-Order Arithmetic}.
\newblock Perspectives in Mathematical Logic. Springer-Verlag, Berlin, 1993.

\bibitem{jerabek:apal04}
E.~Je{\v r}{\'a}bek.
\newblock Dual weak pigeonhole principle, {B}oolean complexity, and
  derandomization.
\newblock {\em Ann. Pure App. Logic}, 129(1--3):1--37, 2004.

\bibitem{krajicek:Bounded-Arithmetic}
J.~Kraj{\'{\i}}{\v{c}}ek.
\newblock {\em Bounded Arithmetic, Propositional Logic, and Complexity Theory},
  volume~60 of {\em Encyclopedia of Mathematics and its Applications}.
\newblock Cambridge University Press, Cambridge, 1995.

\bibitem{krajicek-pudlak:ic98}
J.~Kraj{\'{\i}}{\v{c}}ek and P.~Pudl{\'a}k.
\newblock Some consequences of cryptographical conjectures for {${\rm S}\sp
  1\sb 2$} and {${\rm EF}$}.
\newblock {\em Inform. and Comput.}, 140(1):82--94, 1998.

\bibitem{maciel-etc:wphp}
A.~Maciel, T.~Pitassi, and A.~R. Woods.
\newblock A new proof of the weak pigeonhole principle.
\newblock {\em J. Comput. System Sci.}, 64(4):843--872, 2002.
\newblock Special issue on STOC 2000 (Portland, OR).

\bibitem{pollett:phd}
C.~Pollett.
\newblock {\em Arithmetic Theories with Prenex Normal Form Induction}.
\newblock PhD thesis, University of California, San Diego, 1997.

\bibitem{pollett:apal99}
C.~Pollett.
\newblock Structure and definability in general bounded arithmetic theories.
\newblock {\em Ann. Pure Appl. Logic}, 100(1-3):189--245, 1999.

\bibitem{pollett:mfn-algebras}
C.~Pollett.
\newblock Multifunction algebras and the provability of {${\rm
  PH}\!\downarrow$}.
\newblock {\em Ann. Pure Appl. Logic}, 104(1-3):279--303, 2000.

\bibitem{pollett:aml03}
C.~Pollett.
\newblock On the bounded version of {H}ilbert's tenth problem.
\newblock {\em Arch. Math. Logic}, 42(5):469--488, 2003.

\bibitem{pollett-danner:tcs06}
C.~Pollett and N.~Danner.
\newblock Circuit prinicples and weak pigeonhole variants.
\newblock To appear in \emph{Theoretical Computer Science}.

\end{thebibliography}

\end{document}